\documentclass[twocolumn, tighten]{aastex62}

\usepackage{amsmath}
\usepackage{amssymb}
\usepackage{graphicx}
\usepackage{graphics}
\hypersetup{
     linkcolor  = red,
     urlcolor = magenta,
     filecolor = cyan,
     citecolor  = blue
}

\def\apj{{ ApJ}}
\def\apjl{{ ApJL}}
\def\apjs{{ ApJS}}

\def\aap{{ A\&A}}
\def\aj{{ AJ}}
\def\mnras{{ MNRAS}}

\def\nat {{ Nature}}
\def\pasj{{ PASJ}}

\def\prd{{ Phys. Rev. D}}

\def\pasp{{ PASP}}

\def\mr{\mathrm}

\def\d{\mathrm{d}}

\def\eps{\epsilon}
\def\msun{M_{\rm \odot}}

\graphicspath{{./}{./figures/}}

\shorttitle{The Missing Energy Puzzle of TDEs}
\shortauthors{W. Lu \& P. Kumar}

\begin{document}

\title{On the Missing Energy Puzzle of Tidal Disruption Events}

\correspondingauthor{Wenbin Lu,
  \href{mailto:wenbinlu@caltech.edu}{wenbinlu@caltech.edu}} 

\author{Wenbin Lu}
\affil{TAPIR, Walter Burke Institute for Theoretical Physics, Mail
  Code 350-17, Caltech, Pasadena, CA 91125, USA}
\affil{Department of Astronomy, University of Texas at Austin, Austin,
TX 78712, USA}

\author{Pawan Kumar}
\affiliation{Department of Astronomy, University of Texas at Austin, Austin,
TX 78712, USA}

\begin{abstract}

For the majority of the tidal disruption event (TDE) candidates, the
observed energy in the optical/near-UV bands is of order
$10^{51}\,$erg. We show that this observed energy is
smaller than the minimum bolometric energy for the radiative
inefficient accretion flow model by a
factor of 10{--}100. We argue that this discrepancy is because
the majority of the energy released is in the extreme-UV (EUV)
band and/or in the form of relativistic jets beamed away from the
Earth. The EUV scenario is supported by existing mid-infrared data and
should be further tested by future dust reverberation observations. The
jet scenario is disfavored by the radio observations of ASASSN-14li
but may still be viable for other TDE candidates. We also provide evidence
that, at least for some TDEs, most of the missing energy (in the EUV
and/or in the form of jets) is 
released within a few times the orbital period of the most
tightly bound material $P_{\rm min}$, which means (1) the
circularization of the fall-back stream may occur rapidly and
(2) the luminosity of the accretion flow or the jet power
may not be capped near the Eddington level when the fall-back rate is
super-Eddington. For most other TDEs, this energy-release timescale is
currently not strongly constrained.

\end{abstract}


\section{Introduction}
A tidal disruption event (TDE) occurs when a star is shredded by the
tidal gravity of a supermassive black hole (BH) with mass $M\lesssim
10^8\,\msun$ \citep{1988Natur.333..523R}. To conserve the total
angular momentum, roughly half of the star is ejected to infinity, and
the other half is left in bound elliptical orbits. The bound
materials, after passing 
the apocenters of their orbits, fall back towards the BH and generate
bright multi-wavelength emission. Many TDE 
candidates have been discovered by transient surveys in the 
X-ray \citep[e.g.][]{1999A&A...343..775K, 2002AJ....124.1308D,
  2008A&A...489..543E, 2012A&A...541A.106S}, UV 
\citep[e.g.][]{2008ApJ...676..944G, 2009ApJ...698.1367G}, and optical
bands \citep[e.g.][]{2011ApJ...741...73V, 2012Natur.485..217G,
  2014ApJ...793...38A, 2015ApJ...798...12V, 2016MNRAS.455.2918H}.
As opposed to TDE candidates that typically show a thermal spectrum, a
few events have been characterized by highly variable, non-thermal
(power-law) $\gamma$/X-ray emission, which is likely generated by
relativistic jets pointing towards the Earth \citep{2011Sci...333..203B,
  2011Natur.476..421B, 2011Natur.476..425Z, 2012ApJ...753...77C,
  2015MNRAS.452.4297B}. 

Thermal TDE candidates are usually selected from large samples of 
transient sources by eliminating the by far more common flares from 
variable AGNs (according to the host spectrum) and supernovae (according to
the lightcurve shape, temperature evolution, and the distance to the galactic
center). However, these methods of selection do not 
guarantee the purity of the final TDE sample. For instance, supernovae
from nuclear star clusters
may occur at a rate of
$\sim$$10^{-4}\rm\,galaxy^{-1}\,yr^{-1}$ for 
Milky-Way-like galaxies \citep{2013ApJ...771..118Z}, whereas the
measured rate of the TDE candidates is $10^{-4}${--}$10^{-5}
\rm\,galaxy^{-1}\,yr^{-1}$ \citep{2002AJ....124.1308D, 2014ApJ...792...53V,
  2016MNRAS.455.2918H, 2017arXiv171204936H, 2018ApJ...852...72V}. In
this paper, we make the assumption that 
\textit{at least some} of these candidates are actual TDEs; this is
motivated by the theoretical expectation of TDE rate being
$10^{-4}${--}$10^{-5}\rm\,galaxy^{-1}\,yr^{-1}$
\citep{1999MNRAS.309..447M, 2004ApJ...600..149W, 
  2016MNRAS.455..859S}. This is also supported by the
cutoff in the TDE rate for BH masses above $\sim$$10^8\msun$ due to direct
capture of stars by the event horizon \citep{2018ApJ...852...72V}.

In the generic picture of a TDE, general relativistic apsidal
precession causes the fall-back stream to 
self-intersect, and then the shocks at the collision point convert the
kinetic energy of the upstream gas into thermal energy. The angular momentum of
the stream is redistributed downstream of the shock such that the
orbital eccentricity is reduced and viscous
accretion may proceed efficiently \citep[e.g.][]{1989ApJ...346L..13E,
  1994ApJ...422..508K, 2009ApJ...695..404R, 
  2013ApJ...767...25G, 2013MNRAS.435.1809S, 2015ApJ...804...85S, 
  2016MNRAS.458.4250S, 2016MNRAS.455.2253B,
  2016ApJ...830..125J}. Despite significant progress in recent
works, studying the detailed post-disruption dynamics is 
still challenging due to the three-dimensional nature, the complexity
of the physics (general relativity, self-gravity, radiation, and
magnetohydrodynamics), and the wide range of time/length scales
involved \citep[see][for a review]{2018arXiv180110180S}. An easier and
more robust way of understanding the physics of TDEs is to look at the
global properties such as mass, energy, and angular momentum through
their conservation laws. In the simplest picture,
if half of the star's mass is eventually accreted onto the BH, the
expected energy released is $\sim$$10^{52}${--}$10^{53}\,$erg. However, for 
the majority of the TDE candidates found in recent surveys,
the observed radiation energy in the optical/UV bands is only $\sim$$10^{51}
\,$erg. \textit{The puzzle is: Where is the missing 90{--}99\% of the
  energy?}

This low apparent radiative efficiency led \citet{2015ApJ...806..164P}
to suggest that the optical emission is powered by the
shocks due to stream-stream collision instead of the accretion flow near the
BH. This is because the amount of kinetic energy dissipated at the shocks
is $\sim 10^{51} (r_{\rm I}/10^3r_{\rm 
  g})^{-1}\,$erg, where $r_{\rm I}$ is the intersecting radius and
$r_{\rm g}\equiv GM/c^2$ is the gravitational radius of the BH. However,
\citet{2015ApJ...806..164P} did not consider\footnote{Another potential
issue is that the shock-dissipated energy may not be radiated away
efficiently. In order for the radiation to escape, the
shocked gas needs expand to a much larger 
volume than that occupied by the cold streams and the radiation energy
may be lost in the form of PdV work. The stream-stream collision simulations
by \citet{2016ApJ...830..125J} show that only a few 
percent of the initial kinetic energy is radiated away. 
} the radiation from the
subsequent accretion onto the BH after the stream-stream collision,
which we argue should dominate the energy output from the TDE.
This is because the shocks can only unbind less than half of the gas
in the fall-back stream due to energy requirement. If part of the gas
receives more energy and becomes unbound, the others become more
tightly bound. 

Thus, the majority of the shocked gas must be injected into the
accretion flow and the mass feeding rate to the accretion flow is a
large fraction of that entering the self-intersecting shocks. The
question of the low apparent radiative efficiency \textit{of the
  accretion flow} is still unanswered. There are a number of possible
solutions to this puzzle:

\vspace{-0.1cm}
\begin{itemize}
\item[(1)] The mass-feeding rate to the accretion flow may be highly
super-Eddington at early time, and due to photon trapping and/or mass 
outflow, the luminosity of the escaping radiation may be regulated at a
near-Eddington level. This radiatively inefficient accretion flow model is
widely adopted by many authors \citep{1997ApJ...489..573L,
  1999ApJ...514..180U, 2009MNRAS.400.2070S, 2012ApJ...749...92K,
  2014ApJ...781...82C, 2015MNRAS.453..157P, 2016MNRAS.461..948M}. 

\item[(2)] It has been proposed by \citet{2017MNRAS.467.1426S} that
  the accretion flow may be in the form of a highly eccentric disk
  and the angular momentum exchange among different parts of the disk may
  allow the majority ($\gtrsim$$90\%$) of the mass to fall
  directly into the event horizon without circularization and viscous accretion. 

\item[(3)] Most of the energy released from the accretion flow may be
  in the form of radiation either absorbed along the line of sight
  (e.g. by dust in the host galaxy) or at other wavelengths that have
  not yet been observed.

\item[(4)] Most of the energy released from TDEs may be in
  the form of relativistic jets beamed away from the Earth.
\end{itemize}

We discuss the first scenario in \S 2 and \S 3. Based on the
radiatively inefficient accretion flow model, 
we calculate the \textit{minimum} bolometric energy
output from a TDE and show that it is higher than the observed energy
in the optical/near-UV bands by a factor of 
10{--}100. Thus, this widely adopted model does not solve the energy
efficiency puzzle.

The second scenario requires that (i)
the disk eccentricity stays high for 
$\sim$$10$ orbits ($\sim$$1\,$yr) and (ii) the exchange of angular
momentum among the fluid elements occurs in a way such that only
$\lesssim$$10\%$ of the mass in the accretion flow circularizes and that
the rest plunges into the event horizon of the BH
\citep{2017MNRAS.467.1426S}. Typically, the stream-stream 
collision occurs at a substantial angle \citep[40$^{\rm
  o}${--}160$^{\rm o}$;][]{2015ApJ...812L..39D} and dissipates a large
fraction of the orbital kinetic energy. In addition, the cooling time is
generally comparable to or longer than the orbital time, so the resulting
accretion flow is quasi-spherical \citep{2016MNRAS.461.3760H,
  2016MNRAS.458.4250S} instead of highly elliptical.
This scenario is not discussed in this paper,
because it can only occur in rare cases where the stream-stream
collision angle is small ($\ll$$1\,$rad) and the dissipation of
orbital energy is inefficient.


The third scenario may explain the radiative efficiency puzzle,
considering that the optical/near-UV spectrum of TDEs 
is usually Rayleigh-Jeans-like and that the peak may be in
the extreme-UV (EUV) band. This scenario has not been thoroughly
explored in the literature, because it is difficult to infer the peak
frequency and hence bolometric luminosity from the optical/near-UV
data, considering the strong dependence on the (highly uncertain)
line-of-sight extinction of the host galaxy
\citep[e.g.][]{2009ApJ...698.1367G, 2014ApJ...780...44C}. In \S 4, we
show that this scenario is supported by existing 
dust reverberation observations in the mid-infrared where the
extinction is negligible.

The fourth scenario, which involves relativistic off-axis jets, will
be discussed in \S5. We show that this scenario is ruled out by radio
observations of ASASSN-14li \citep{2014MNRAS.445.3263H,
  2016Sci...351...62V, 2016ApJ...819L..25A}, but it still remains
viable for the other thermal TDEs with only sparse radio upper
limits.

In \S 6, we briefly discuss the implications of including the missing
energy on the physics of circularization and accretion processes in
TDEs. Throughout the paper, the convention of $Q 
= 10^nQ_n$ and cgs units are used.
\section{Mass-feeding rate to the accretion flow}

We consider a star of mass $M_* = m_*\msun$ and radius $R_* =
r_*R_{\odot}$ interacting with a BH of mass $M = 10^6M_6\msun$.
The tidal disruption radius is expressed in units of the BH's gravitational
radius $r_{\rm g}\equiv GM/c^2$ as
\begin{equation}
  \label{eq:12}
  {r_{\rm T}\over r_{\rm g}} \simeq {R_*\over r_{\rm g}} \left( {M
      \over M_*} \right)^{1/3} \simeq 47\, M_6^{-2/3} m_*^{-1/3} r_*.
\end{equation}
When the star first enters the radius $r_{\rm T}$, the
tidal gravity of the BH causes a spread of specific energy across 
the star $-\Delta\eps \leq \eps \leq \Delta \eps$ and
\citep{2013MNRAS.435.1809S} 
\begin{equation}
  \label{eq:95}
  \Delta \epsilon =\eta_{\eps} \frac{GMR_*}{r_{\rm T}^2}\simeq
  2.1\times10^{-4} c^2 \, \eta_{\eps} M_6^{1/3} m_*^{2/3} r_*^{-1},
\end{equation}
where $\eta_{\eps}$ is a dimensionless number of order unity
(depending on the stellar density profile and the detailed disruption
process). Bound materials have negative specific energies
$-\Delta \epsilon \leq\epsilon<0$ and the leading edge ($\epsilon =
-\Delta \epsilon$) has the minimum orbital period 
\begin{equation}
  \label{eq:24}
  P_{\rm min} \simeq (41\,\mathrm{d})\,
\eta_{\eps}^{-3/2} M_6^{1/2} m_*^{-1} r_*^{3/2}.
\end{equation}

\begin{figure}
  \centering
\includegraphics[width = 0.48\textwidth,
  height=0.27\textheight]{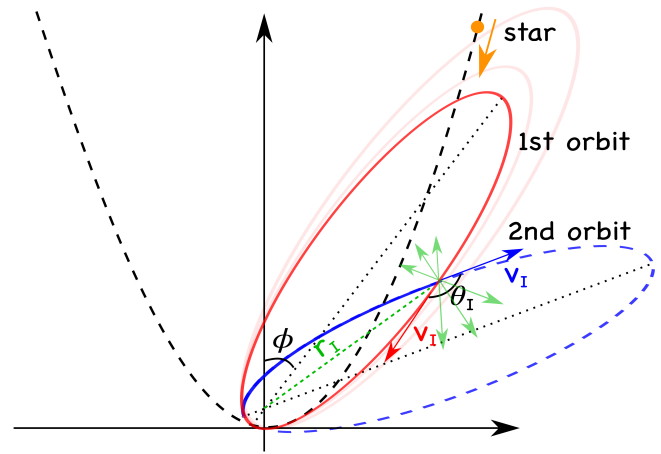}
\caption{A schematic picture of a tidal disruption event. The star was
  initially in a parabolic orbit (dashed black 
  curve). After the tidal 
  disruption, the bound materials are in highly eccentric elliptical
  orbits of different semimajor axes (the most bound orbit is
  shown in red) but have the same apsidal
  precession angle $\phi$ per orbital cycle. Materials in their second
  orbit collide with those in the first orbit with velocity
  $v_{\rm I}$ at radius $r_{\rm I}$ from the BH, and the intersecting
  angle is $\theta_{\rm I}$.
Then, the post-shock gas expands rapidly into a thick torus wrapping
around the BH.
}\label{fig:orbit}
\end{figure}

As shown in Fig. (\ref{fig:orbit}), the bound stream
self-intersects due to apsidal\footnote{We ignore the
Lense-Thirring precession arising from the BH's spin, which 
may otherwise deflect the stream out of the initial orbital
plane. If the stream has a small enough cross section
\citep[the evolution of which is controlled by the competition among
tidal shear, self-gravity, gas and magnetic pressure, etc.;
see][]{1994ApJ...422..508K, 2016MNRAS.455.3612C}, such a deflection may  
prevent stream-stream intersection during the second orbit and hence
circularization can be significantly delayed
\citep{2015ApJ...809..166G}. As shown in \S 3, our conclusion on the
radiation energy discrepancy will be even stronger because the
mass-feeding rate to the disk may be reduced and the radiative
efficiency of the accretion flow should
be closer to that of the standard thin disk
\citep{1973A&A....24..337S}.} precession. Since the intersecting
radius $r_{\rm I}$ is typically less than the apocenter radius of the
most bound orbit, the time it takes for a fluid element to move from the
beginning of the second orbit to the intersecting point is insensitive
to the specific energy. Thus, the mass-flow rate to the
intersecting point is given by
\begin{equation}
  \label{eq:6}
  \begin{split}
\dot{M} \simeq {\d M\over \d P} = {\d M\over \d \epsilon} {\d
 \epsilon\over \d  P} = {\d M\over \d \epsilon} {(2\pi GM)^{2/3} \over
3} P^{-5/3},
  \end{split}
\end{equation}
which depends on the (uncertain) specific energy distribution $\d M/\d
\epsilon$.

\begin{figure*}
  \centering
\includegraphics[width = 0.9\textwidth,
  height=0.35\textheight]{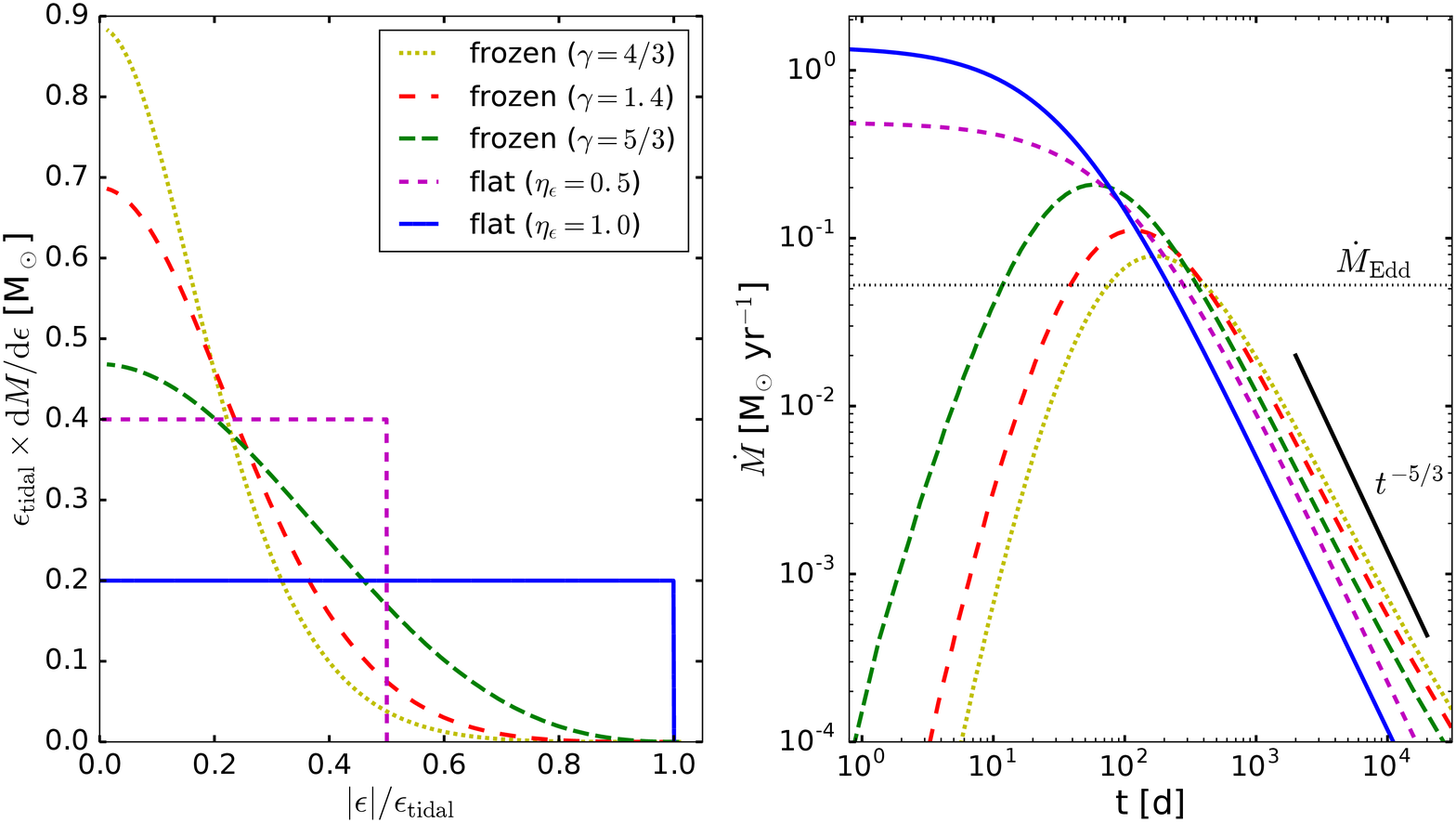}
\caption{{\it Left panel:} the yellow, red, and green curves show the
  distributions of specific energy when the 
  star first enters $r_{\rm T}$, under the assumption that it keeps
  its spherical shape and original density profile. We assume a
  polytropic equation of state and solve the
  Lane-Emden equation under different polytropic indexes $\gamma=4/3,
  1.4$, and $5/3$. The specific energy is normalized by
   the work per unit mass done by tidal forces $\eps_{\rm
     tidal}\equiv GMR/r_{\rm T}^2$. The
  magenta and blue curves show a flat distribution of specific energy in
  ($-\Delta \epsilon$, 0), where $\Delta\epsilon = \eta_{\eps}
  \eps_{\rm tidal}$. We use BH mass $M = 10^6\rm\ M_{\rm \odot}$, stellar mass
  $0.4\msun$ and stellar radius $0.48R_{\rm \odot}$ for this
  figure.
\newline {\it Right panel:} mass fall-back rate for different cases in
the left panel. We define $t=0$ to be when the most tightly bound
material (with $\eps = -\Delta \eps$) reaches the intersection point.
At late time, the fall-back rate goes as $t^{-5/3}$ since $\d M/\d
\epsilon$ is asymptotically flat. 
The horizontal dashed line shows the Eddington accretion rate
$\dot{M}_{\rm Edd}=5.3\times10^{-2}\,\rm 
\msun\,yr^{-1}$ according to eq. (\ref{eq:16}) with thin-disk radiation
efficiency $\eta = 0.05$.
}\label{fig:feeding}
\end{figure*}

In the following, we consider two simple models of specific energy
distribution: (i) in the first (``frozen'')
case \citep[e.g.][]{2009MNRAS.392..332L, 2013MNRAS.435.1809S,
  2016MNRAS.455.3612C}, the distribution of specific energy is assumed
to be ``frozen'' when the star first enters the tidal radius $r_{\rm
  T}$ and each fluid element evolves ballistically (ignoring self-gravity, internal
pressure, and shocks), and (ii) in the second (``flat'') case
\citep[e.g.][]{1989ApJ...346L..13E, 1989IAUS..136..543P}, $\d M/\d
\epsilon$ is assumed to be independent of $\eps$ in the range
$-\Delta \epsilon <\epsilon<\Delta \epsilon$, i.e.,
\begin{equation}
  \label{eq:18}
{\d M\over \d\epsilon}  = {M_*\over 2\Delta\eps} = 2.3\times10^{3}
{\msun \over c^2}\, \eta_{\eps}^{-1} M_6^{-1/3} m_*^{1/3} r_*.
\end{equation}
We note that the specific energy distributions
given by Newtonian and relativistic numerical simulations
generally lie between these two extreme cases \citep[e.g.][]{2013ApJ...767...25G,
  2014PhRvD..90f4020C, 2015ApJ...808L..11C, 2017MNRAS.469.4483T}. We
also define $t=0$ as when the most tightly bound material (with $\eps
= -\Delta \eps$) reaches the intersecting point.

Since the self-intersecting shock can only unbind a small fraction of
the gas in the fall-back stream, the mass enters the accretion
flow at a rate roughly equal to that entering the shock.\footnote{Alternatively,
  one can assume that a fraction $f_{\rm b}\lesssim 1$ of the
  fall-back gas joins the accretion flow. This factor can be easily
  included by replacing the thin-disk efficiency $\eta$ (see
  eq. \ref{eq:16} for definition) by $f_{\rm
    b}\eta$ throughout this section, and our conclusion stays
  unchanged.} In the ``frozen'' case, when the star first enters the
tidal radius $r_{\rm T}$, we assume that its spherical shape and original density profile
are preserved and that all fluid elements are moving at the same
velocity $v_0 = \sqrt{2GM/r_{\rm T}}$. Then, the specific orbital
energy of each fluid element is assumed to be frozen until the stream
self-intersects. For a polytropic equation of state $P\propto
\rho^{\gamma}$, we solve the Lane-Emden equation for the density
profile given by the total stellar 
mass $M_*$ and radius $R_*$. We consider three different polytropic
indexes $\gamma=4/3, 1.4, 5/3$ (the star is more centrally
concentrated for lower $\gamma$).

In the ``flat'' case, the mass-feeding rate to the accretion flow is
\begin{equation}
  \label{eq:17}
  \dot{M}(t) \simeq 3.0\mr{\,\msun\,yr^{-1}} {\eta_{\eps}^{3/2}
  m_*^2 \over M_6^{1/2} r_*^{3/2}} \left({t + P_{\rm min}\over P_{\rm
        min}}\right)^{-5/3},
\end{equation}
which shows the canonical $t^{-5/3}$ behavior at $t\gg P_{\rm min}$:
  \begin{equation}
    \label{eq:9}
    \dot{M}(t)\simeq
    7.8\times10^{-2}\mr{\,\msun\,yr^{-1}} 
    {M_6^{1/3}m_*^{1/3}r_*\over \eta_\eps} \left(t\over
      \mr{yr}\right)^{-5/3}. 
  \end{equation}
We note that the fall-back rates calculated via grid
  \citep{2013ApJ...767...25G} or smoothed-particle
  hydrodynamical \citep{2015ApJ...808L..11C} 
  simulations for \{$M_6=1, m_* = 1, r_*=1, \gamma=5/3$\} are
  quite similar with $\dot{M}(t=1\mr{\,yr}) \simeq 9\times10^{-2}
  \mr{\,\msun\,yr^{-1}}$, which is close to our ``flat'' case
  if $\eta_\eps\simeq 0.9$.

In Fig. (\ref{fig:feeding}), we show the distribution of specific
energy (left panel) and the mass fall-back rates (right
panel) for the ``frozen'' and ``flat'' cases.  We use $M_* =
0.4\msun$, $R_* = 0.48R_{\odot}$, and $M = 10^6\msun$. At sufficiently
late time $t\gg P_{\rm min}$, the mass feeding rate in all cases approaches the
characteristic $t^{-5/3}$ power-law, because the specific energy
distribution is asymptotically flat, $\d M/\d \eps\approx\rm
constant$. The peak mass fall-back rate is much smaller in the
``frozen'' cases, because more mass is concentrated near $\eps\approx
0$. 

We define the Eddington accretion rate as 
\begin{equation}
  \label{eq:16}
  \dot{M}_{\rm Edd} \equiv {L_{\rm Edd}\over \eta c^2} =
(2.6\times10^{-2}\,\mr{\msun\,yr^{-1}}) \,M_6 \eta_{-1}^{-1}
\end{equation}
where $L_{\rm Edd} =1.5\times10^{44} M_6\rm \,erg\, 
s^{-1}$ is the Eddington luminosity for solar metallicity with Thomson
scattering opacity and $\eta = 0.1\eta_{-1}$ is the radiative
efficiency\footnote{Note that $\eta = 1 -
\sqrt{1 - 2r_{\rm g}/(3r_{\rm ISCO})}$ ranges from 0.038 (for
a retrograde orbit around an extreme Kerr BH with spin $-1$) to 0.42
(for a prograde orbit around an extreme Kerr BH with spin $1$), where
$r_{\rm ISCO}$ is the radius of the innermost stable circular
orbit. For a zero-spin Schwarzschild BH, we have $\eta = 0.057$.
} for a standard thin disk \citep{1973A&A....24..337S}. In the
``flat'' case, the ratio between the peak mass-feeding rate and the
Eddington accretion rate is
\begin{equation}
  \label{eq:19}
  {\dot{M}_{\rm peak}\over \dot{M}_{\rm Edd}}\simeq 114\,
  \eta_{\eps}^{3/2}\eta_{-1} M_6^{-3/2} m_*^2 r_*^{-3/2},
\end{equation}
which becomes less than unity for BH mass $M\gtrsim 2.3\times10^7
\msun\, \eta_{\eps}\eta_{-1}^{2/3} m_*^{4/3} r_*^{-1}$. On the right
panel of Fig. (\ref{fig:feeding}), we mark
the Eddington accretion rate for the (conservative) efficiency $\eta =
0.05$ with a horizontal dashed
line. In all the cases considered in Fig. (\ref{fig:feeding}), the mass
fall-back rates exceed the Eddington accretion rate for $\sim$$1\,$yr.

\begin{figure*}
  \centering
\includegraphics[width = 0.9\textwidth,
  height=0.37\textheight]{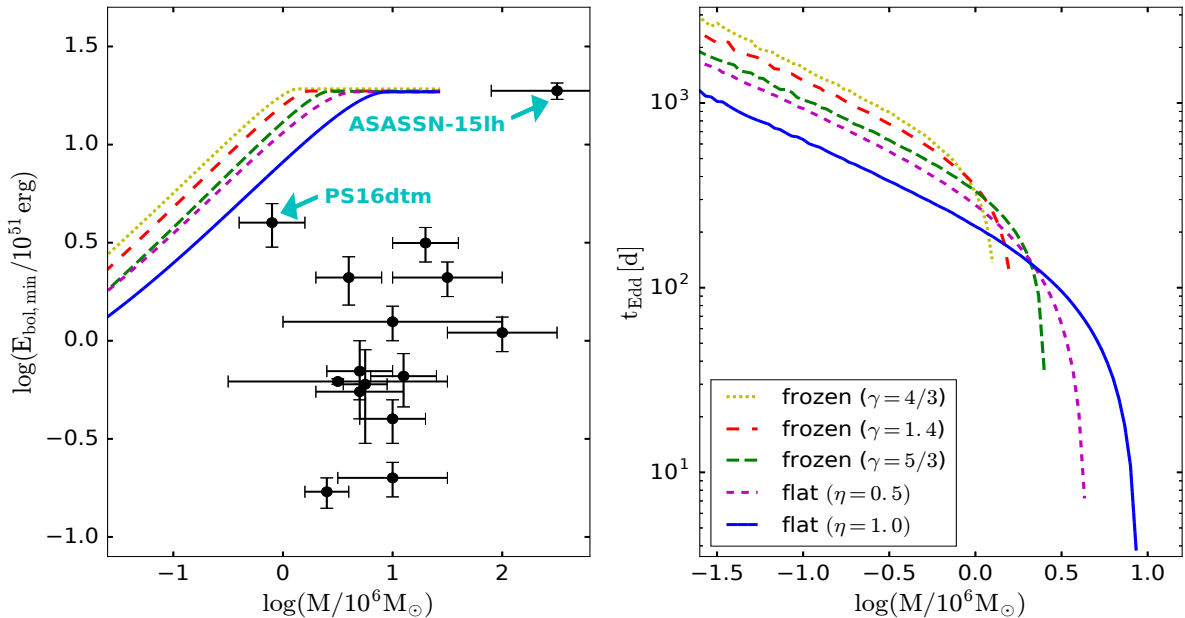}
\caption{\textit{Left panel:} the minimum bolometric energy
released corresponding to the mass-feeding rate to the accretion flow
on the right panel of Fig. (\ref{fig:feeding}). When the feeding rate
is super-Eddington ($\dot{M}\geq L_{\rm Edd}/\eta c^2$), we adopt a
minimum bolometric 
luminosity $L_{\rm bol,min}\simeq L_{\rm Edd}$; when the feeding rate is
sub-Eddington ($\dot{M}< L_{\rm Edd}/\eta c^2$), the accretion flow
settles into a thin disk, and the bolometric luminosity is $L_{\rm bol}
\simeq \eta \dot{M} c^2$. We use a conservative thin-disk radiation
efficiency parameter $\eta = 0.05$ for this figure. We also plot the observed
radiation energy in the optical/near-UV bands from the TDE candidates
D3-13, D1-9, 
D23H-1, TDE1, TDE2, Dougie, PTF09ge, iPTF16axa, PS1-10jh, PS1-11af,
PS16dtm, and ASASSN-14ae, -14li, -15oi, -15lh. We find that the
observed radiation energy in the optical/near-UV bands (in the form of
a blackbody component with temperature 1{--}3$\times10^4\,$K) is
typically a factor of 10{--}100 smaller than the minimum bolometric energy
expected in the generic TDE model, except for two cases, ASASSN-15lh
and PS16dtm (where the discrepancies are less than a factor of $\sim$2).
\newline
\textit{Right panel:} the duration $t_{\rm Edd}$ over which the mass
feeding rate to the accretion flow exceeds the Eddington accretion
rate. We find that for BH masses $\lesssim$ a few$\times10^6\msun$, 
super-Eddington mass-feeding lasts for $t_{\rm Edd}$$\gtrsim$a
few$\times10^2${--}$10^3$ days.
}\label{fig:Ebol_tEdd}
\end{figure*}

\section{Minimum bolometric energy}

Super-Eddington BH accretion flows ($\dot{M}\gg
\dot{M}_{\rm Edd}$) are generally believed to be radiatively
inefficient compared to sub-Eddington thin disks, because (i)
radiation can be trapped by the 
inflowing matter and advected into the BH \citep{1979MNRAS.187..237B},
and (ii) the majority of the 
accreting mass may become unbound due to energy injection along
with the outward transport of angular momentum
\citep{1994ApJ...428L..13N, 1999MNRAS.303L...1B,
  2004MNRAS.349...68B}. As a result, the bolometric radiative 
luminosity may be regulated at the Eddington
luminosity times a logarithmic factor of order unity, as shown in
recent multidimensional simulations of super-Eddington 
BH accretion disks \citep{2011ApJ...736....2O, 2014MNRAS.441.3177M,
  2016MNRAS.456.3929S}. We note that vertical advection of
radiation due to magnetic buoyancy may transport radiation energy
faster than diffusion, which may allow the accretion disk to radiate at highly
super-Eddington luminosities \citep[e.g.][]{2011ApJ...733..110B,
  2014ApJ...796..106J}, although the radiation may suffer
from adiabatic loss if the disk also launches a highly optically thick
wind. Another possibility for higher radiative efficiency is that the
accumulation of magnetic flux near the BH event horizon leads to a
magnetically-arrested disk (magnetic fields become dynamically
important), where radiation escapes along the low-density channel 
created by the Poynting flux of rotating field lines
\citep{2003PASJ...55L..69N, 2015MNRAS.454L...6M}.

In this section, we calculate the \textit{minimum} bolometric energy
from a TDE, according to the mass-feeding rates to the
accretion flow obtained in the previous section. Our calculation is
based on two conservative assumptions: (i) when the mass-feeding rate
to the accretion flow exceeds the Eddington accretion rate, the
angle-integrated bolometric luminosity is \textit{at least} $L_{\rm
  Edd}$, and (ii) when the mass-feeding rate drops below the Eddington
accretion rate, the bolometric emission from the disk tracks the
mass-feeding rate and the  
luminosity becomes $\eta \dot{M} c^2$ as in the standard thin-disk
model, i.e., 
\begin{equation}
  \label{eq:22}
  \begin{cases}
    L_{\rm bol}(t)\gtrsim L_{\rm Edd}\ &\mbox{if }\dot{M}(t) \geq \dot{M}_{\rm Edd};\\
    L_{\rm bol}(t)\simeq \eta \dot{M} c^2\ &\mbox{if }\dot{M}(t) < \dot{M}_{\rm Edd}.
  \end{cases}
\end{equation}
The total bolometric radiation energy is given by
\begin{equation}
  \label{eq:1}
  E_{\rm bol} \equiv \int_0^{\infty} L_{\rm bol}(t)\d t.
\end{equation}
Note that, if the accretion is significantly delayed (e.g. due to
slow circularization and/or viscous spreading) with respect to the
mass-injection rate given by eq. (\ref{eq:6}), the peak accretion rate
is reduced and the total bolometric radiation energy will be
\textit{higher}\footnote{General relativistic
  hydrodynamical simulations by \citet{2015ApJ...804...85S} showed that
  the circularization process after the initial stream-stream
  collision may take up to $\sim$$10P_{\rm min}$ ($\sim$$1\,$yr) and
  hence the accretion process may be significantly lengthened at a
  reduced accretion rate. The viscous evolution of the accretion flow
  with a time-dependent mass feeding has been studied by
  \citet{1990ApJ...351...38C} and \citet{2014ApJ...784...87S}. TDEs with a
  prolonged accretion phase of $\sim$$1\,$yr and slow variability
  may have been  missed in current optical/UV surveys (T. Holoien 2018,
  private communication).} (since the radiation efficiency of the accretion flow
is closer to that of a standard thin disk). 

We also denote the duration of super-Eddington accretion as $t_{\rm
 Edd}$ (and $t_{\rm Edd}=0$ if $\dot{M}_{\rm peak} < \dot{M}_{\rm
 Edd}$). In the ``flat'' specific energy distribution case, if
$\dot{M}_{\rm peak}\gg \dot{M}_{\rm Edd}$, $t_{\rm
 Edd}$ can be solved from $\dot{M}(t_{\rm Edd}) = \dot{M}_{\rm Edd}$
(combining eqs. \ref{eq:17} and \ref{eq:16}),
\begin{equation}
  \label{eq:2}
  t_{\rm Edd} \simeq (700\mr{\,d})\, \eta_{\eps}^{-3/5}
  \eta_{-1}^{3/5} M_6^{-2/5} m_*^{1/5} r_*^{3/5},
\end{equation}
Then, the minimum bolometric energy is
\begin{equation}
  \label{eq:23}
  \begin{split}
      E_{\rm bol}  &\gtrsim L_{\rm Edd} t_{\rm Edd} + \int_{t_{\rm
      Edd}}^{\infty} L_{\rm Edd} \left({t\over t_{\rm
        Edd}}\right)^{-5/3} \d t \simeq 
  {5\over 2} L_{\rm Edd} t_{\rm Edd}\\
  & \simeq (2.3\times
  10^{52}\mr{\,erg})\, \eta_{\eps}^{-3/5} \eta_{-1}^{3/5} M_6^{3/5}
  m_*^{1/5} r_*^{3/5},
  \end{split}
\end{equation}
where we have used $\dot{M}(t)\propto t^{-5/3}$ when $t\gg P_{\rm
  min}$. Note that eqs. (\ref{eq:2}) and (\ref{eq:23}) are only valid when
$\dot{M}_{\rm peak}\gg \dot{M}_{\rm Edd}$, i.e. for BH masses
$M\lesssim (1.1\times 10^7\mr{\,\msun})\, 
\eta_{\eps} \eta_{-1}^{2/3} m_*^{4/3} r_*^{-1}$; for larger BH
masses, the super-Eddington duration $t_{\rm Edd}$ quickly drops to
zero, and the bolometric energy released $E_{\rm bol}$ approaches $\eta
M_* c^2/2 \simeq (9\times10^{52}\mr{\,erg})\, \eta_{-1}m_*$.

On the left panel of Fig. (\ref{fig:Ebol_tEdd}), we show the minimum
bolometric energy released corresponding to the mass-feeding rates
shown on the right panel of Fig. (\ref{fig:feeding}). We use thin-disk
radiative efficiency parameter $\eta = 0.05$ for this figure. We 
also plot the observed radiation energy in the optical/near-UV bands
from the TDE candidates found by \textit{GALEX} \citep[D3-13, D1-9,
D23H-1;][]{2006ApJ...653L..25G, 2008ApJ...676..944G,
  2009ApJ...698.1367G}, 
\textit{SDSS} \citep[TDE1, TDE2;][]{2011ApJ...741...73V}, \textit{ROTSE}
\citep[Dougie;][]{2015ApJ...798...12V}, \textit{PTF} \citep[PTF09ge,
iPTF16axa;][]{2014ApJ...793...38A, 2017ApJ...842...29H}, \textit{Pan-STARRS}
\citep[PS1-10jh, PS1-11af, PS16dtm;][]
{2012Natur.485..217G, 2014ApJ...780...44C, 2017ApJ...843..106B}, 
and \textit{ASASSN} \citep[-14ae, -14li, -15oi, -15lh;][]{2014MNRAS.445.3263H,
  2016MNRAS.455.2918H, 2016MNRAS.463.3813H,
  2016Sci...351..257D}. Only Galactic extinction has been
corrected for these events. Our sample is selected from the open TDE 
catalog (\href{http://tde.space}{http://tde.space})
based on the good quality of the
optical/near-UV photometry in these sources. We find that the observed
radiation energy in the optical/near-UV bands, typically in the form
of a blackbody component with temperature (1{--}3)$\times10^4\,$K, is a
factor of 10{--}100 smaller than the minimum bolometric energy expected
in the generic TDE model. This conclusion is not sensitive to the
different fallback-rate models.
We note that ASASSN-15lh and PS16dtm are the only two cases with
observed radiation energies near the theoretical
expectation to within a factor of $\sim$2. It is still
debated whether these two sources are TDEs or superluminous supernovae 
\citep[see][]{2015MNRAS.454.3311M, 2016Sci...351..257D, 
  2016NatAs...1E...2L, 2017ApJ...843..106B}. We also note that the
missing energy is not observed in the X-ray band (0.2{--}10$\,$keV)
because the ratios between the X-ray and optical/near-UV luminosities
for these TDE candidates are at most of order unity
\citep{2017ApJ...838..149A}.

It can be seen from eq. (\ref{eq:23}) that this energy discrepancy
cannot be reconciled by varying the thin-disk radiative efficiency
$\eta$$\in$(0.038, 0.42), specific energy distribution parameter
$\eta_{\eps}$$\in$$(\sim$0.3, $\sim$1), stellar mass $m_*\gtrsim 0.1$ 
(for a main-sequence star),  and stellar radius $r_*\gtrsim
0.2$.

We note that highly centrally concentrated stars (e.g. the Sun)
  are only fully disrupted for deeply penetrating orbits, while for
  the more general situation, the total
  fall-back mass is smaller than $M_*/2$
  \citep{2017A&A...600A.124M}. However,
  it is unlikely that \textit{the majority of} the observed
  TDEs are partial disruptions where the star only loses
  $\ll$$0.1\,\msun$ of mass, because the bright optical emission
  requires a minimum amount of 
mass to reprocess the EUV or soft X-ray emission from the accretion
flow \citep{2015ApJ...806..164P, 2016MNRAS.461..948M,
  2016ApJ...827....3R}. In the following, we show that this minimum
reprocessing layer constrains the total fall-back mass to be more than
$\sim$$0.1\msun$ for most TDEs in our sample.

The blackbody-like optical/near-UV spectrum of a TDE gives
the emitting area
\begin{equation}
  \label{eq:11}
  A = L_{\rm BB}/\sigma T^4\simeq
(1.8\times10^{30}\mr{\, cm^2})\,
L_{\rm BB,44}T_{4.5}^{-4},
\end{equation}
where $\sigma$ is the Stefan-Boltzmann
constant, $L_{\rm BB}$ is the luminosity of the optical/near-UV
component, and $T$ is the photospheric temperature. 
If the emitting surface covers a solid angle of $\Omega$ ($\leq
4\pi$),  the corresponding photospheric radius is
\begin{equation}
  \label{eq:13}
  r_{\rm  ph}\simeq (3.7\times10^{14}\mr{\,cm})\, L_{\rm
  BB,44}^{1/2} T_{4.5}^{-2}\sqrt{4\pi/\Omega},
\end{equation}
The location of the
photosphere is marked by the effective absorption optical depth
$\tau\simeq \rho\kappa r_{\rm ph}\simeq 1$, where the effective
opacity is given by $\kappa =
\sqrt{\kappa_{\rm abs}\kappa_{\rm s}}$ in the scattering-dominated
regime, and $\kappa_{\rm abs}$ and $\kappa_{\rm s}$ are the opacities
for absorption and electron 
scattering respectively. Thus, the mass density near
the photosphere is given by
\begin{equation}
  \label{eq:14}
  \rho\simeq (\kappa r_{\rm ph})^{-1}\simeq
(10^{-13.5}\mr{\,g\,cm^{-3}})\, \kappa_{-1}^{-1}r_{\rm ph,14.5}^{-1},
\end{equation}
where we have used a fiducial effective opacity of $\kappa =
0.1\kappa_{-1}\rm\,cm^2\, g^{-1}$ motivated by the fact that
bound-free and free-free opacity
$\kappa_{\rm abs}\ll \kappa_{\rm s}\simeq 0.34 \rm\,cm^2\, g^{-1}$
(for solar abundance) at densities $\rho\lesssim  
10^{-13}\rm\, g\, cm^{-3}$ and temperatures $T\gtrsim 10^4\rm\,K$ (a
rough scaling is $\kappa_{\rm abs}\propto \rho T^{-3.5}$). The true
opacity depends on gas metallicity and the detailed ionization state
near the photosphere \citep[see a discussion by][]{2016ApJ...827....3R}.

On the other hand, since the fall-back materials do not have
enough angular momentum to circularize at radii
$r_{\rm ph}\gtrsim10^{14}\rm\,cm$, the widths of the 
(H$\alpha$ and/or He$\,$II)  emission lines cannot be sufficiently
broadened by gas rotation. Instead, they suggest that the gas in the
line-forming region near the 
photosphere is expanding outwards at typical speeds
$v=10^9v_9\rm\,cm\,s^{-1}$. To sustain the bright optical emission for
a duration of $\Delta t$ (typically $\sim$$30$ days), the reprocessing
gas needs to have a minimum mass\footnote{A speculative ``TDE impostor'' idea
  is the sudden accretion of $\sim$$0.01\msun$ following stellar
  collisions due to consecutive extreme mass ratio inspirals
  \citep[EMRIs,][]{2017ApJ...844...75M}. In such cases, the peak
  accretion luminosity can reach $\sim$$10^{44}\rm\,erg\,s^{-1}$ but
  mainly in the EUV or soft X-ray bands. To match the
  observed optical/near-UV luminosity of
  $\sim$$10^{44}\rm\,erg\,s^{-1}$, this model relies on the
  reprocessing by a fossil disk (with mass $\sim$$0.1\msun$) at distances 
  10{--}100$\,$AU from the BH. The disk must exist for
  $\gtrsim$$10^4\,$yr to witness the stellar collision and hence is
  geometrically thin. It is unclear whether
  such a thin disk is capable of reprocessing a large fraction (order
  unity) of the EUV or soft X-ray emission from the TDE impostor.
}
\begin{equation}
  \label{eq:15}
  M_{\rm rep}\gtrsim \rho A v\Delta t\simeq (0.06\msun) {L_{\rm
    BB,44}^{1/2} \over T_{4.5}^{2} \kappa_{-1}} \sqrt{\Omega\over 4\pi} 
  {v_9\Delta t\over 30\mr{\,d}}.
\end{equation}
To launch this outflow through local dissipation of
orbital energy \citep[e.g. via stream-stream
collision as shown in][]{2014ApJ...796..106J}, the total amount of fall-back
(bound) mass needed is $M_{\rm b}\gtrsim 2M_{\rm rep}\gtrsim
0.1\msun$.

Another possible solution to the energy discrepancy is that the BH
masses in these TDEs are actually $\lesssim$$10^5\,\msun$. For some of
the TDEs in our sample, BH masses lower (by up to a factor of 10) than
the values reported in the discovery papers are indeed found by
\citet{2017MNRAS.471.1694W} from the velocity dispersion of the host
galaxies using high-resolution spectroscopy. On the right panel of
Fig. (\ref{fig:Ebol_tEdd}), we show 
the duration $t_{\rm Edd}$ over which the mass-feeding rate to the accretion flow
exceeds the Eddington accretion rate. We can see
that, if the luminosity is indeed capped near the Eddington luminosity
of $\lesssim$$10^5\msun$ BHs, then the radiation energy should be released
on a timescale of $\gtrsim$$10^3$ days (eq. \ref{eq:2}), which is 
inconsistent with the observed rise and decay time in the TDEs in
our sample. Additionally, many TDE candidates had peak
optical/near-UV luminosities of order $10^{44}\rm\,erg\,s^{-1}$, which
is much higher than $L_{\rm Edd}$ for BH masses $\lesssim10^{5}\msun$.

\section{The extreme-UV scenario}

In the previous section, we show that 10{--}100 times
the observed energy may have escaped our detection. For the TDEs in
our sample, the optical/near-UV spectrum is usually
Rayleigh-Jeans-like, so most of the missing energy may be in the EUV
band (there are stringent upper limits in the X-ray band). This
scenario is also motivated by the temperature associated with an
Eddington luminosity emerging at the tidal disruption radius
\citep{1999ApJ...514..180U} 
\begin{equation}
  \label{eq:32}
  \begin{split}
      kT \simeq k\left({L_{\rm Edd}\over 4\pi r_{\rm T}^2
      \sigma_{\rm SB}}\right)^{1/4}
 \sim (20\mr{\,eV})\, M_6^{1/12} r_*^{-1/2} m_*^{1/6},
  \end{split}
\end{equation}
where $k$ is the Boltzmann constant and $\sigma_{\rm SB}$ is the
Stefan-Boltzmann constant. This EUV emission may be directly observable
for high-redshift TDEs, provided that photons near the spectral peak
do not suffer from strong 
absorption (by e.g. neutral hydrogen) in the host galaxy.

In this section, we discuss another indirect way of constraining the
total UV-optical luminosity with dust reverberation mapping in the
mid-infrared. If the
gaseous medium at the galactic center is dusty, a fraction of the UV photons 
will be absorbed by dust and re-radiated in the mid-infrared
wavelengths $\lambda \sim3$-$10\rm\, \mu m$ \citep{2016MNRAS.458..575L}. In
the optically thin limit, due to high UV fluxes, dust particles will
sublime within the critical radius \citep{2000ApJ...537..796W}
\begin{equation}
  \label{eq:25}
  R_{\rm sub} \sim (0.2\mr{\,pc})\, L_{\rm UV,45}^{1/2} (T_{\rm
    sub}/1800\mr{\,K})^{-2.5} a_{-5}^{-1/2},
\end{equation}
where $L_{\rm UV} = 10^{45}L_{\rm UV,45}\rm\,erg\,s^{-1}$
is the \textit{total} UV-optical luminosity of the TDE, $T_{\rm sub}$
is the sublimation temperature, and $a = 0.1a_{-5}\rm\,\mu m$ is the
grain radius. The mid-infrared emission from the 
surviving dust particles at radius $R\sim R_{\rm sub}$ lasts for a
light-crossing timescale of the radiating dust shell
\begin{equation}
  \label{eq:21}
    t_{\rm IR}\sim {R_{\rm sub} \over c} \simeq (0.6\,\mr{yr})\, L_{\rm
      UV,45}^{1/2} \left({T_{\rm sub}\over 1800\mr{\,K}}\right)^{-2.5} a_{-5}^{-1/2}.
\end{equation}
Therefore, we can monitor the mid-infrared emission from a sample of
newly discovered TDEs on a cadence of a few months and then use the
measured duration $t_{\rm IR}$ to infer the total UV-optical luminosity $L_{\rm
UV}$.

We note that eqs. (\ref{eq:25}) and (\ref{eq:21}) only capture the
qualitative behavior of dust reverberation mapping. More detailed
calculations are needed to extract the total UV-optical luminosity from
mid-infrared observations. The grain temperature at a certain distance
from the UV source depends on the infrared emissivity, which is 
uncertain at high temperatures $\gtrsim$$1000\,$K
\citep{1984ApJ...285...89D}. The sublimation temperature $T_{\rm sub}$
depends on the size (distribution) and composition of dust particles
\citep{1989ApJ...345..230G}. More importantly, the mid-infrared 
duration $t_{\rm IR}$ depends on the spatial distribution of dust near
the galactic center as well as on the observational wavelength
(the emission spectrum from grains of a certain temperature is not a blackbody
and colder dust particles emit at longer wavelengths). For instance,
different dust distribution geometries (e.g. torus, patchy clouds)
could cause $t_{\rm IR}$ to vary by a factor of a few, so the
UV-optical luminosity may have uncertainty of an order of magnitude
($L_{\rm UV}\propto t_{\rm IR}^2$). More detailed calculations taking
these uncertain factors into account are left for future works.

Currently, a few TDEs have
been observed by the \textit{Wide-field Infrared Survey Explorer}
at 3.6 and 4.5$\,\rm \mu m$ \citep{2016ApJ...828L..14J,
  2016ApJ...829...19V, 2017ApJ...841L...8D}. PTF09ge had mid-infrared
luminosity of $L_{\rm IR}\sim 10^{42}\rm\,erg\,s^{-1}$ lasting for
$t_{\rm IR}\sim 1\,$yr \citep{2016ApJ...829...19V}. This implies an EUV
luminosity of $L_{\rm UV}\sim 10^{45}\rm\,erg\,s^{-1}$ (but note the large
uncertainties mentioned above), which is a factor of
$\sim$$10$ higher than the optical/near-UV luminosity. If the duration
of the EUV emission equals to that of the optical emission, then the
total energy released from PTF09ge may be of order $10^{52}\,$erg. Future
multi-wavelength mid-infrared observations by \textit{JWST} will enable
detailed modeling of the dust properties/spatial distribution, and
hence we can measure the total UV-optical 
energy released from TDEs more accurately. This method of reverberation
mapping, commonly used in the AGN community
\citep[e.g.][]{1993PASP..105..247P}, can be used to constrain the 
total energy reservoir of TDEs and may provide a solution to the
energy efficiency puzzle. 


\section{The Jet Scenario}

In this section, we discuss the possibility that
$10^{52}${--}$10^{53}\,$erg of energy is carried away by
narrowly beamed relativistic jets. This is
motivated by the powerful non-thermal $\gamma$/X-ray emission from
Swift J1644+57 \citep{2011Sci...333..203B, 2011Natur.476..421B}, Swift
J2058+05 \citep{2012ApJ...753...77C}, and perhaps a 
third one, Swift J1112+82 \citep{2015MNRAS.452.4297B}. The isotropic
equivalent energies released in the X-ray band in Swift 
J1644+57 and J2058+05 are $E_{X}^{\rm (iso)}\simeq
5\times10^{53}\,$erg and $9\times10^{53}\,$erg, respectively. The
total energy carried by the jet is given by $E_{\rm j}=E_{\rm X}f_{\rm
  b,X} f_{\rm bol}^{-1} f_{\rm 
  r}^{-1}$, where $f_{\rm b,X}$ is the beaming factor
of the X-ray emission, $f_{\rm bol}^{-1}$ is the bolometric correction
(to account for $\gamma$-rays), and $f_{\rm r}$ is the jet
radiative efficiency. Despite large uncertainties in these
factors, we can see that the jet may indeed carry up to
$\sim$$10^{53}\,$erg of energy in these cases (e.g. if $f_{\rm
  b,X}\sim 0.02$ and $f_{\rm bol}^{-1}f_{\rm  r}^{-1}\sim 10$ for
Swift J1644+57).

There are a number of arguments against this jet scenario for the
majority of TDEs. First, only about $10\%$ of AGNs are radio-loud with
relativistic jets \citep[e.g.][]{1995ApJ...438...62W}, but this argument may be
weak if the properties of the TDE accretion flow differ qualitatively
from those in AGNs. We also note that the jet power in radio-loud AGNs
is on the same order as the disk luminosity
\citep[e.g.][]{2014Natur.515..376G}, so the EUV scenario and the jet
scenario may operate together.

Second, the rate of events like Swift J1644+57 is estimated to be
$\sim$$10^{-9}\rm\,galaxy^{-1}\,yr^{-1}$, if one 
assumes that Swift BAT is sensitive to all events within its field of
view ($\sim$$4\pi/7$) in the past $\sim$10$\,$yr up 
to redshift $\sim$1 \citep{2011Natur.476..421B, 2012ApJ...753...77C,
  2015MNRAS.452.4297B}. This extremely small 
rate is 4 orders of magnitude lower than the observationally inferred
TDE rate of $\sim$$10^{-5}\rm\,galaxy^{-1}\,yr^{-1}$ and it requires a
combination of a small beaming factor and that only a small
fraction of TDEs produce relativistic jets. However, the fact that all
three jetted TDEs were discovered within three months of the same year
raises the question of incompleteness. Additionally, relativistic jets from
TDEs may release the majority of their radiation in the X-ray band,
and they do not trigger Swift BAT in the $\gamma$-ray band. Thus, the
observed non-thermal TDEs may be the tip-of-an-iceberg cases of a much
larger population of jetted events, so this argument is also a weak one.

Third, powerful radio emission as seen in Swift J1644+57 is expected
to be observable even for off-axis observers at sufficiently late times
$t\gtrsim1\,$yr when the jet decelerates 
to non-relativistic speeds \citep{2011MNRAS.416.2102G}. However, the
flux at a given frequency seen by an off-axis  
observer depends sensitively on the poorly 
known circum-nuclear medium (CNM) density profile, the fraction of energy
in magnetic fields in 
the shocked region ($\eps_{\rm B}$), and the angular structure of the jet
\citep{1998ApJ...497L..17S, 2003ApJ...591.1075K}. Many of the
thermal TDEs in our sample only had (typically single-epoch) upper
limits\footnote{Similar upper limits have been reported by \citet{2013ApJ...763...84B}
for a sample of X-ray-selected TDE candidates.} at $5\,$GHz at
$\sim$1{--}10 yr after the optical peak \citep{2013A&A...552A...5V}. 
\citet{2017MNRAS.464.2481G} explored a wide range of CNM densities and
used these radio upper limits to constrain the jet
energies to be $E_{\rm j}\lesssim10^{52}${--}$10^{53}\,$erg (mostly near
the higher end; see their Table 2). In their modeling, they
kept fixed the magnetic energy fraction ($\eps_{\rm B}=0.002$) and the
angular structure of the jet \citep[a fast narrow core with Lorentz factor 10
surrounded by a slow wide sheath with Lorentz factor 2, as
in][]{2015MNRAS.450.2824M}. Including the uncertainties in these two
aspects (especially lower $\eps_{\rm B}$) may make their constraints
even weaker.

The most constraining case is ASASSN-14li (luminosity distance $D_{\rm
  L}\approx 90\,$Mpc) where the 
radio luminosity is many orders of magnitude lower and decays rapidly
since the first detection \citep[at $\sim$1$\,$month after the optical
peak;][]{2016Sci...351...62V}. In the following, we show that a
powerful jet with energy $E_{\rm j}\sim 10^{52}${--}$10^{53}\,$erg
is inconsistent with multi-wavelength observations from 1.45 to
24.5$\,$GHz at late  
time $t\gtrsim 150\,$d \citep{2016ApJ...819L..25A}.

At sufficiently
late times, the shock front expands in a nearly spherical manner
at non-relativistic speed $\beta c$ and radius $r\simeq \beta c t$
from the BH. We assume the CNM density profile $n(r)$ to be sufficiently
smooth (e.g. a power-law function), so the total number of
protons (or electrons) in the shocked region is
\begin{equation}
  \label{eq:5}
  N_{\rm tot} \simeq {E_{\rm j} \over m_{\rm p} \beta^2 c^2/2} \simeq
  {4\pi\over 3} (\beta c t)^3 n,
\end{equation}
which gives $n\simeq (54\mr{\,cm^{-3}})\,E_{\rm
  j,52}\beta^{-5}(t/150\mr{\,d})^{-3}$ at radius $r$. The magnetic field
strength in the shocked region is
\begin{equation}
  \label{eq:3}
  B \simeq (16\pi \epsilon_{\rm B} n m_{\rm p} \beta^2
  c^2)^{1/2}\simeq 
  (2.7\times10^{-3}\mr{\,G})\, n^{1/2} \beta \eps_{\rm B,-4}^{1/2}.
\end{equation}
The spectral peak is at the synchrotron self-absorption
frequency $\nu_{\rm a}$, at which the flux density $F_{\nu_{\rm a}}$
is given by
\begin{equation}
  \label{eq:4}
  4\pi D_{\rm L}^2 F_{\nu_{\rm a}}\simeq {2\nu_{\rm a}^2\over
    c^2} \gamma_{\rm a}m_{\rm e} c^2 4\pi^2 r^2, 
\end{equation}
where $\gamma_{\rm a} = (4\pi m_{\rm e}c \nu_{\rm a}/3eB)^{1/2}$ is
the electron Lorentz factor corresponding to synchrotron frequency
$\nu_{\rm a}$. We combine eqs. (\ref{eq:5}), (\ref{eq:3}) and
(\ref{eq:4}), and obtain
\begin{equation}
  \label{eq:7}
  F_{\nu_{\rm a}} \simeq (125\mr{\,mJy})\, \beta^{11\over 4}
  \nu_{\rm a,9}^{5\over 2} (t/150\mr{\,d})^{11\over 4} E_{\rm
    j,52}^{-{1\over 4}}\eps_{\rm B,-4}^{-{1\over 4}}.
\end{equation}
Therefore, in the off-axis jet scenario, the shock expansion speed can
be calculated from the three measurables $\nu_{\rm a}$, $F_{\nu_{\rm
a}}$, $t$ by
\begin{equation}
  \label{eq:8}
  \beta\simeq 0.17\, (F_{\nu_{\rm a}}/\mr{mJy})^{4\over 11} \nu_{\rm
    a,9}^{-{11\over 10}} (t/150\mr{\,d})^{-1} E_{\rm
    j,52}^{1\over 11}\eps_{\rm B,-4}^{1\over 11}.
\end{equation}
The measured peak frequency $\nu_{\rm a}$ and flux density
$F_{\nu_{\rm a}}$ at $t=(143,\,
207,\, 246,\, 304,\, 381)\,$d by \citet{2016ApJ...819L..25A} give
expansion speeds $\beta\simeq (2.2, 
\,2.6,\, 2.4,\, 2.9,\, 2.8) \times10^{-2}(E_{\rm j,52}\eps_{\rm
  B,-4})^{1/11}$. A similar conclusion of nearly constant expansion speed is
drawn by \citet{2016ApJ...827..127K}, under the assumption that
magnetic field and electrons are in energy equipartition (although we
do not make this assumption). This nearly free expansion of the
``synchrotron photosphere'' at a sub-relativistic speed seen in the
radio band is inconsistent with the deceleration of a blast wave in
the Sedov regime, and hence the relativistic off-axis jet scenario with
$E_{\rm j}\sim 10^{52}${--}$10^{53}\,$erg is ruled out for
ASASSN-14li.

We conclude that the jet scenario, where most of the
energy released from the TDE accretion flow is in the form of narrowly
beamed relativistic jets, is ruled out for ASASSN-14li, but it is
still a viable solution to the energy efficiency puzzle for the other
thermal TDEs. This is mostly because single-epoch
upper limits are by far less constraining than multi-wavelength and
multi-epoch detections.

\section{Discussion}

In previous sections, we have shown that 90\%{--}99\% of the energy
released from the accretion flow is not detected for the majority of 
TDEs. In the following, we briefly discuss the implications of 
the missing energy on the physics of circularization and accretion in TDEs. 

We argue that, at least for some TDEs, the missing energy (in the EUV and/or in the
form of relativistic jets) may be released within a few times the orbital
period of the most bound material $P_{\rm min}$ (see 
eq. \ref{eq:24}). This is based on the following two aspects of
observations. (i) The X-ray emission from jetted TDEs
Swift J1644+57 \citep{2011Sci...333..203B, 2011Natur.476..421B} and
Swift J2058+05 \citep{2012ApJ...753...77C} show a plateau
lasting for two weeks and then a power-law decay roughly as $L_{\rm
X}\propto t^{-5/3}$, roughly tracking the fall-back rate. (ii) From
the reverberation geometry, one can see that the rise or decay time of
the dust mid-infrared emission provides an upper limit for the
duration of the EUV emission. The first and second mid-infrared
detections of ASASSN-14li was 35 and 200 rest-frame days
(respectively) after discovery or the 
optical peak \citep{2016ApJ...828L..14J}, with flux dropping by a
factor of 5 between these two epochs. We can see that the duration of
the EUV emission must be $\lesssim$$100\,$d (the most conservative
case is that the mid-infrared flux rose in first 100$\,$d and then
immediately dropped in the second 100$\,$d).

For many other TDEs, the timescale over which the
missing energy is released is unconstrained and hence the EUV or jet
luminosity is uncertain. For instance, the constraints on the
EUV duration from PTF09ge and PTF09axc \citep{2016ApJ...829...19V}
are rather weak, because the mid-infrared observations were made much
later. PTF09ge had the first and second detections at
200 and 360 rest-frame days after the optical peak, with roughly equal
fluxes between the two epochs. Then in the most conservative scenario,
the duration of the EUV emission may be up to 280$\,$d, which is a
weak constraint. We also note that the X-ray emission (associated with
the accretion process) from ASASSN-14li had decay time of
$\sim$$30\,$d \citep{2015Natur.526..542M}, while the X-ray emission
from ASASSN-15oi steadily rises from 
$L_{\rm X}\sim 5\times 10^{41}\rm\,erg\,s^{-1}$ to
$5\times10^{42}\rm\,erg\,s^{-1}$ over one year after discovery
\citep{2017ApJ...851L..47G, 2018arXiv180400006H}. The X-ray lightcurve
from ASASSN-15lh is also qualitatively similar as -15oi
\citep{2017ApJ...836...25M}. Since only a small
fraction of the total energy released is in the X-ray band ($L_{\rm
  X}\ll L_{\rm Edd}$), the diverse X-ray lightcurves could be due to
variable efficiency of high-energy emission and hence do not provide
strong constraints on the circularization or disk-formation
timescale.

\section{Summary}

For TDE candidates found by recent
surveys, the observed radiation energy in the optical/near-UV bands is only
$\sim$$10^{51}\,$erg, which is much smaller than the total amount of energy
generated $\sim$$10^{52}${--}$10^{53}\,$erg if roughly half of the star's mass
is accreted onto the BH. In this paper, we show that this energy
discrepancy cannot be explained by the radiatively inefficient
accretion flow model at super-Eddington accretion rate. We calculated
the minimum bolometric 
energy output from a TDE, based on the assumption that the minimum
radiative luminosity from the accretion flow is $\mbox{min}(L_{\rm Edd}, \eta
\dot{M}c^2)$, where $\eta$ is the radiative efficiency of a standard
thin disk and $\dot{M}$ is the mass-feeding rate to the accretion
flow. The minimum bolometric energy from our calculation is higher
than the observed energy in the optical/near-UV bands by a factor of
10{--}100. Therefore, we argue that the missing energy may be in the
unobserved EUV band and/or in the form of relativistic jets. 

The EUV scenario is supported by the observations that the
optical/near-UV spectrum of TDEs is close to the Rayleigh-Jeans law
and that the luminosity in the X-ray band is at most comparable to
that in the optical/near-UV (for thermal TDEs). This
scenario is also supported by existing mid-infrared dust reverberation
observations, although the quality of currently existing data is
relatively low in terms of time and wavelength coverages. Future dust
reverberation mapping will not only provide a measurement of the total
EUV luminosity from the TDE but also constrain the composition and
spatial distribution of dust particles near non-active galactic
nuclei.

The jet scenario is disfavored by a number of
indirect (and weak) arguments. The strongest constraint is from
ASASSN-14li, where relativistic off-axis jets with $E_{\rm
  j}\sim 10^{52}${--}$10^{53}\,$erg have been ruled out by
multi-wavelength radio observations. It is currently unclear whether the
radio emission from ASASSN-14li is representative of all the other
thermal TDEs, so the jet scenario still
remains a viable solution to the energy efficiency puzzle for some
TDEs. Future wide field-of-view X-ray transient surveys may directly
probe the rate of jetted TDEs and provide better constraints on this
scenario. 

We also provide evidences that, at least for some TDEs, most of the
missing energy (in the EUV and/or in the form of jets) is released
within a few times the orbital period of the most tightly bound
material $P_{\rm min}$, which means (1) the circularization of the fall-back
stream may occur rapidly, and (2) the luminosity of the accretion flow
or the jet power may not be capped near the Eddington level when the
fall-back rate is super-Eddington. For most other TDEs, this energy-release
timescale is currently not strongly constrained. In the future, more
realistic numerical simulations are needed to understand the detailed
circularization and accretion processes. 

\section{acknowledgments}
We thank Sjoert van Velzen and James Guillochon for reading the
manuscript and providing valuable comments. The comments from an
anonymous referee have improved the clarity of the paper. We also
thank Tom Holoien, Julian Krolik, Scott Tremaine, Jeremy Goodman, and Brian
Metzger for useful discussions. W.L. was supported by the David
Alan Benfield Memorial Fellowship at the University of Texas at
Austin, and the David and Ellen Lee Fellowship at California Institute
of Technology.

\end{document}